\documentclass{moriond}
\usepackage{wrapfig}
\usepackage{cite}
\usepackage{amsmath}

\def\Journal#1#2#3#4{{#1} {\bf #2}, #3 (#4)}

\def\PRD{{\em Phys. Rev.} D}

\def\be{\begin{equation}}
\def\ee{\end{equation}}
\def\bea{\begin{eqnarray}}
\def\eea{\end{eqnarray}}

\usepackage{amsmath}
\newcommand{\Photo}{}

\begin{document}
\vspace*{4cm}
\title{CORRECTIONS TO MODELS OF B-PHYSICS DECAY ANOMALIES}

\author{ NUD\v ZEIM SELIMOVI\'C }

\address{Physik-Institut, Universit\"at Z\"urich, CH-8057 Z\"urich, Switzerland}

\maketitle\abstracts{Motivated by the recent anomalies observed in semileptonic decays of B-mesons, the class of models denoted as "4321" is introduced and analysed. The importance of calculating next-to-leading-order (NLO) effects in performing precise compatibility tests of data with 4321 predictions is discussed. On the one hand, the coupling of the $U_1 \sim (\textbf{3},\textbf{1})_{2/3}$ leptoquark appearing in the spectrum to the third-generation fermions is large, giving rise to sizable NLO corrections in the relevant semileptonic operators. On the other hand, the new source of flavor violation is introduced with the vector-like fermions, resulting in the processes whose relevance can only be quantified at the loop order.}

\section{Introduction}

It has been almost a decade now since the first deviations from the Standard Model (SM) predictions were observed in semileptonic decays of B-mesons \cite{Lees:2013uzd}. The measurements by various collaborations that followed lead to a coherent pattern of the SM deviations, which are now known as "B anomalies". The hints collected so far are grouped into two categories: i) deviations from $\tau/\mu$ and $\tau/e$ universality in $b \to c\ell\bar{\nu}$ charged currents, and ii) deviations from $\mu/e$ universality in $b \to s\ell^+\ell^-$ neutral currents. As shown in the recent study \cite{Cornella:2021sby}, the combined significance of deviations is at the level of $4\sigma$ for both charged and neutral current observables.\newline
\indent In order to provide a combined explanation, it was recognized that ingredients of the New Physics (NP) should be: i) an approximate $U(2)^5$ flavour symmetry implying dominant couplings of the new sector to the third generation fermions \cite{Barbieri:2015yvd}, and ii) a TeV-scale vector leptoquark, $U_1 \sim (\textbf{3},\textbf{1})_{2/3}$, as a particularly successful mediator. The different UV-complete models incorporating the two ingredients~\cite{DiLuzio:2017vat,DiLuzio:2018zxy,Fuentes-Martin:2020bnh,Cornella:2019hct} share the TeV-scale dynamics which is always characterized  by the gauge group $\mathcal{G}_{4321} \equiv SU(4) \times SU(3) \times SU(2) \times U(1)$ acting in a family non-universal way. As a general feature, these models collectively denoted as "4321", contain new gauge fields acquiring mass in the $\mathcal{G}_{4321} \to \mathcal{G}_{SM}$ spontaneous symmetry breaking. Moreover, realistic realisations involve the vector-like fermions which induce couplings between the new $SU(4)$ vectors and the light SM generations. \newline
\indent In this talk I want to report some of the most important results concerning the NLO effects associated with the new fields, calculated in a series of papers \cite{Fuentes-Martin:2019ign, Fuentes-Martin:2020luw, Fuentes-Martin:2020hvc}. The motivation comes from the expectation that these corrections are sizable since coupling of the leptoquark to SM fermions must be large $(\mathcal{O}(3))$ in order to explain B-physics data, while being consistent with collider searches. Additionally, vector-like fermions introduce a new source of flavor violation which results in the flavour-changing neutral current (FCNC) amplitudes absent at tree-level. 

\section{The 4321 models}

%%%%%%%%%%%%%%%%%%%%%%%%%%%%%%%%%%%%%%%%%%%%%%%%%%

The 4321 models are based on the $\mathcal{G}_{4321} \equiv SU(4)\times SU(3)^\prime\times SU(2)_L\times U(1)_X$ gauge symmetry, with the corresponding gauge fields denoted by $H_\mu^{A=1,\dots,15}$, $C_\mu^{a=1,\dots 8}$, $W_\mu^{1,2,3}$ and $B^\prime_\mu$, and the gauge couplings $g_4$, $g_3$, $g_2$ and $g_1$, respectively. The SM gauge group corresponds to the $\mathcal{G}_{4321}$ subgroup $SU(3)_c\times U(1)_Y\equiv[SU(4)\times SU(3)^\prime\times U(1)_X]_{\rm diag}$, with $SU(2)_L$ being the SM one. It is useful to define the mixing angles $\theta_{1,3}$, relating the 4321 gauge couplings to the SM ones
\begin{equation}
    g_s=g_4\,\sin\theta_3=g_3\, \cos\theta_3\,,\quad
    g_Y=\sqrt{\tfrac{3}{2}}\,g_4\, \sin\theta_1=g_1\, \cos\theta_1\,,
\end{equation}
with $g_s$ and $g_Y$ denoting the $SU(3)_c$ and $U(1)_Y$ gauge couplings. In terms of the 4321 gauge eigenstates, the SM gluon, $G_\mu^a$, and hypercharge gauge boson, $B_\mu$, can be written as
\begin{equation}
    G_\mu^a=\cos\theta_3\,C_\mu^a+\sin\theta_3\,H_\mu^a\,,\quad
    B_\mu=\cos\theta_1\,B_\mu^\prime+\sin\theta_1\,H_\mu^{15}\,,
\end{equation}
while the new massive vectors $G_\mu^{\prime\,a}\,,Z'_\mu\,,U_\mu^{1,2,3}$ read 
\begin{eqnarray}
    G_\mu^{\prime\,a}&=-\sin\theta_3\,C_\mu^a+\cos\theta_3\,H_\mu^a\,,\qquad
    Z_\mu^\prime=-\sin\theta_1\,B_\mu^\prime+\cos\theta_1\,H_\mu^{15}\,,\nonumber\\
    &U_\mu^{1,2,3}=\frac{1}{\sqrt{2}}(H_\mu^{9,11,13}-iH_\mu^{10,12,14})\,.
\end{eqnarray}
\begin{wraptable}{r}{0.5\textwidth}
\vspace{-0.8cm}
\caption{Minimal matter content.}
\vspace{0.2cm}
\centering
\setlength{\tabcolsep}{2.5pt}
\renewcommand{\arraystretch}{1.1}
\begin{tabular}{|c|c|c|c|c|c|}
\hline
Field & $SU(4)$ & $SU(3)^\prime$ &  $SU(2)_L$  & $U(1)_X$ \\
\hline
\hline 
$\psi_L$ & $\mathbf{4}$  & $\mathbf{1}$  & $\mathbf{2}$  &  0 \\
$\psi^\pm_R$ & $\mathbf{4}$ & $\mathbf{1}$   & $\mathbf{1}$  &  $\pm 1/2$   \\ 
$q_L^{\prime\,i}$ & $\mathbf{1}$  & $\mathbf{3}$ & $\mathbf{2}$  &  $1/6$   \\[2pt]
$u_R^i$ & $\mathbf{1}$  & $\mathbf{3}$ & $\mathbf{1}$  &  $2/3$   \\[2pt]
$d_R^i$ & $\mathbf{1}$  & $\mathbf{3}$ & $\mathbf{1}$  &  $-1/3$   \\[2pt]
$\ell_L^{\prime\,i}$ & $\mathbf{1}$  & $\mathbf{1}$ & $\mathbf{2}$  &  $-1/2$   \\[2pt]
$e_R^i$ & $\mathbf{1}$  & $\mathbf{1}$ & $\mathbf{1}$  &  $-1$   \\
\hline
\hline
%$H$ & $\mathbf{1}$ & $\mathbf{1}$ & $\mathbf{2}$ & $1/2$\\
$\Omega_3$ & $\mathbf{\bar 4}$ & $\mathbf{3}$ & $\mathbf{1}$ & $1/6$\\
$\Omega_1$ & $\mathbf{\bar 4}$ & $\mathbf{1}$ & $\mathbf{1}$ & $-1/2$\\
\hline
\end{tabular}
\label{tab:minimal_content}
\end{wraptable}
In most 4321 models, the $\mathcal{G}_{4321} \to \mathcal{G}_{SM}$ breaking is triggered by the vacuum expectation values of two scalar fields transforming in the antifundamental of $SU(4)$, $\Omega_1$ and  $\Omega_3$, singlet and triplet under $SU(3)^\prime$, respectively.\footnote{An additional scalar field, transforming in the adjoint of $SU(4)$ and singlet under the rest, is often introduced is some 4321 models~\cite{DiLuzio:2017vat,Cornella:2019hct}.} The would-be SM fermions are charged non-universally under the 4321 gauge group, as summarised in Table~\ref{tab:minimal_content}. Here $i=1,2$, $\psi_L\equiv(q_L^{\prime 3}\; \ell_L^{\prime 3})$, $\psi_R^+\equiv(u_R^3\; \nu_R^ 3)$ and $\psi_R^-\equiv(d_R^3\; e_R^3)$. The prime in the fields indicates that these are not mass eigenstates.\newline

\noindent \begin{wraptable}{r}{0.5\textwidth}
\vspace{-0.89cm}
\begin{center}
\begin{minipage}{0.45\textwidth}
\caption{Additional fermion content. Here $\chi_L=(Q_L^\prime\;L_L^\prime)$ and $\chi_R=(Q_R\;L_R)$. The prime in the $\chi_L$ components indicates that these are not mass eigenstates.}\vspace{0.2cm}
\end{minipage}

\setlength{\tabcolsep}{2pt}
\renewcommand{\arraystretch}{1}
\begin{tabular}{|c|c|c|c|c|c|c|}
\hline
Model & Field & $SU(4)$ & $SU(3)^\prime$ &  $SU(2)_L$  & $U(1)_X$ \\
\hline
& $\chi_L$ & $\mathbf{4}$  & $\mathbf{1}$  & $\mathbf{2}$  &  0 \\ 
I & $Q_R$ & $\mathbf{1}$  & $\mathbf{3}$  & $\mathbf{2}$  &  $1/6$ \\ 
& $L_R$ & $\mathbf{1}$  & $\mathbf{1}$ &  $\mathbf{2}$  & $-1/2$ \\
\hline 
\hline 
II& $\chi_{L,R}$ & $\mathbf{4}$  & $\mathbf{1}$  & $\mathbf{2}$  &  0\\
\hline
\end{tabular}
\end{center}
\label{tab:VLcontent}
\end{wraptable}In order to introduce the couplings of $SU(4)$ vectors to the light generations, and 2-3 CKM matrix elements, we add to the minimal content one family of left-handed fermions, transforming in the fundamental of $SU(4)$ and $SU(2)_L$, and one family of right-handed partners. The massive fermions are vector-like under the SM gauge group, therefore, the right-handed partners should transform in the fundamental of $SU(2)_L$, but there is freedom in the $SU(4)\times SU(4)^\prime(\supset SU(3)')$ transformations. As shown in Table~\ref{tab:VLcontent}, we consider two possible realisations: Model I discussed in \cite{Fuentes-Martin:2020bnh}, and Model II in \cite{Cornella:2019hct}. In both cases, having two $SU(4)$ charged fermions, $\psi_L$ and $\chi_L$, leads to a new flavor symmetry that we denote $U(2)_\xi$, with $\xi_L = (\psi_L, \chi_L)^T$. This symmetry is broken by the fermion masses, causing mixing among $\psi_L$ and $\chi_L$, and also the $SU(4)$-singlet fermions, $q_L^\prime$ and $\ell_L^\prime$. Whatever the source of $SU(4)$ breaking is, the mass terms read
\begin{equation}\label{eq:MassMix}
\mathcal{L}_{\rm mass}  = 
\bar\Psi_L^{q\,\prime}\, M_q\,Q_R+\bar\Psi_L^{\ell\,\prime}\, M_\ell\,L_R\,,
\end{equation}
with the left-handed fermions arranged into the flavor vectors $ \Psi^{q\,\prime}_L=(q_L^{\prime\, 2},q_L^{\prime\, 3},Q_L^\prime)^T$ and $ \Psi^{\ell\,\prime}_L=(\ell_L^{\prime\, 2},\ell_L^{\prime\, 3},L_L^\prime)^T$,
and $M_{q,\ell}$ being the $3$-dimensional mass vectors. Without loss of generality, these mass vectors can be written as 
\begin{equation}
M_{q,\ell}=\tilde W_{q,\ell}\,O_{q,\ell}\begin{pmatrix} 0\  \ 0\ \ m_{Q,L} \end{pmatrix}^T\,,
\end{equation}
where $m_{Q,L}$ are the vector-like fermion masses. Here, the $3\times3$ unitary matrices $\tilde W_{q,\ell}$ parametrize the mixing among $SU(4)$ states, while the $3\times3$ orthogonal matrices $O_{q,\ell}$ parametrize the mixing among different $SU(4)$ representations. Explicitly, these matrices read
\begin{equation}\label{eq:Omatrices}
\tilde W_{q,\ell}=
    \begin{pmatrix}
    1 & 0\\
    0 & W_{q,\ell}
    \end{pmatrix}
    \,,\quad O_{q,\ell}=
\begin{pmatrix}
\cos\theta_{Q,L} & 0 & \sin\theta_{Q,L}\\
0 & 1 & 0\\
-\sin\theta_{Q,L} & 0 & \cos\theta_{Q,L}\\
\end{pmatrix}
\,,
\end{equation}
where $\theta_{Q,L}$ are the mixing angles, and $W_{q,\ell}$ being $2\times 2$ unitary matrices. To better understand the origin of the flavor mixing matrices, it is convenient to follow the analogy with the SM. Here, the $SU(4)$ group puts quarks and leptons into a multiplet, which will be misaligned after the $SU(4)$ breaking, similar to the up-down quark misalignment inside the $SU(2)_L$ multiplet of the SM. This misalignment will cause the quark-lepton flavour mixing which is parametrised by the physical combination of $W_q$ and $W_\ell$ matrices, $W_q^\dagger W_\ell=W$, appearing in the $U_1$ interaction, and can be viewed as a generalization of the CKM matrix to $SU(4)$ or quark-lepton space. In this sense, the $U_1$ leptoquark is analogous to the SM $W$, while the $Z^\prime, G^\prime$ are analogous to the SM $Z$. These interactions can be written in the $SU(4)$ basis, or in the quark ($\mathcal{Q}_L^i$) and lepton ($\mathcal{L}_L^i$) components of $\xi^i_L$, that in the mass-eigenstate
basis are given by
\begin{equation}\label{eq:SU4basis}
\begin{pmatrix}
0\\[2pt]
\mathcal{Q}_L^1\\[2pt]
\mathcal{Q}_L^2
\end{pmatrix}
=
P_{23}\,O_q
\begin{pmatrix}
q_L^2\\[2pt]
q_L^3\\[2pt]
Q_L
\end{pmatrix}
\,,\quad
\begin{pmatrix}
0\\[2pt]
\mathcal{L}_L^1\\[2pt]
\mathcal{L}_L^2
\end{pmatrix}
=
P_{23}\,O_\ell
\begin{pmatrix}
\ell_L^2\\[2pt]
\ell_L^3\\[2pt]
L_L
\end{pmatrix}
\,,
\end{equation}
with $P_{23}=\rm diag (0,1,1)$ a projector onto $SU(4)$ states. The interactions read ($i=1,2$)
\begin{equation}
\mathcal{L}\supset\frac{g_4}{\sqrt{2}}\,(\mathcal{\bar Q}_L^i\,W_{ij}\,{U\mkern-13.5mu/}\, \mathcal{L}_L^j+{\rm h.c.})+\frac{g_4}{2\sqrt{6}}\,\left(\mathcal{\bar Q}_L^i{Z^\prime\mkern-16.5mu/}\,\, \mathcal{Q}_L^i-3\,\mathcal{\bar L}_L^i{Z^\prime\mkern-16.5mu/}\,\, \mathcal{L}_L^i\right)+g_4\,\mathcal{\bar Q}_L^i{G^\prime\mkern-16.5mu/}\,\,^a\, T^a \mathcal{Q}_L^i\,.
\end{equation}
As in the SM, FCNCs proportional to the $W$ matrix are generated at the loop level. Computing these contributions to $\Delta F=1$ semileptonic and dipole operators, as well as $\Delta F=2$ hadronic and leptonic amplitudes, represents one of the main results of this work.

\section{NLO results}

\subsection{Gauge sector}

In this subsection I report the main results concerning the relation between low- and high-energy observables beyond the tree level. In particular, we evaluate the NLO corrections in two largest couplings, $g_4(m_U^2)$ and $g_s(m_U^2)$, to the Wilson coefficients of the dimension-six semileptonic operators involving SM third-generation fermions, with $m_U$ being the $U_1$ leptoquark mass. The following normalization is adopted
\begin{equation}
 \mathcal L = -\frac{g_4^2}{2m_U^2}\sum_k \mathcal C_k(\mu) \mathcal O_k\,,
\end{equation}
with the most relevant operators being those generated by the $U_1$ tree-level exchange
\begin{equation}
 \mathcal{O}_{LL}^U = (\bar{\ell}'^3_L \gamma^\mu {q'}_L^3)(\bar{q}'^3_L \gamma_\mu {\ell'}_L^3)\,, \quad
 \mathcal{O}_{LR}^{U} = -2(\bar {\ell}'^3_L {e}'^3_R)(\bar{d}'^3_R {q}'^3_L) + \mathrm{h.c.}\,.
 \end{equation}
Firstly, we calculated one-loop corrections to the corresponding 4-fermion amplitudes in the full theory, employing the dimensional regularisation with $\overline{MS}$ renormalisation scheme as the most convenient because of the infrared singularities and the precise knowledge of $g_s(\mu)$. Proceeding this way, an additional step of expressing the (unphysical) coupling $g_4$ in the Lagrangian in terms of some high-energy physical observable is needed. To this end, we chose the inclusive leptoquark width, and used it to define the corresponding physical coupling. Secondly, the  matching  between the full  and  the  effective  theory was performed. %As a result, we obtain the finite matching corrections proportional to $\alpha_4(m_U^2)$ and $\alpha_s(m_U^2)$ due to the new gauge dynamics. 
The main result is that both NP and QCD corrections are positive, together reaching $20\%$ ($40\%$) increase in $\mathcal{C}_{\rm LL}$ ($\mathcal{C}_{\rm LR}$), implying enhanced $U_1$ contribution at low-energy using the fixed high-energy inputs \cite{Fuentes-Martin:2019ign,Fuentes-Martin:2020luw}. This is an important phenomenological consequence meaning that all collider bounds dominated by
the on-shell production of the new states will be weaker at fixed low-energy contribution once the quantum corrections are taken into account.
\subsection{Flavour changing neutral currents}
The 4321 models are exposed to relevant constraints originating from FCNC processes not present at tree-level, but receiving contributions from $U_1$ loops proportional to the $W$ matrix. As notable examples, I want to assess the UV sensitivity of $B_s - \bar{B}_s$ mixing and $B\to K^{(*)}\nu\bar{\nu}$ transition. In the case of $B_s - \bar{B}_s$ mixing, the one-loop contribution to the Wilson coefficient parametrising this process is
\begin{equation}
    \mathcal{C}_{bs}^{\rm NLO} = \frac{g_4^2}{2 m_U^2} \frac{\alpha_4}{4\pi}\left(\sin\theta_Q\cos^2\theta_L W_{12}W_{22}^*\right)^2 \left(\frac{m_L^2}{4m_U^2} + \mathcal{O}\left(\frac{m_L^4}{m_U^4}\right)\right)\,,
\end{equation}
and is completely controlled by the vector-like lepton mass, $m_L$, for fixed charged-current anomaly.
This can be used to extract the upper bound on $m_L$, and in combination with direct searches squeeze the parameter space for these particles, putting them in the $\mathcal{O}(1\ {\rm TeV})$ ballpark. Similarly, we performed the first complete analysis of the $U_1$ impact in $\mathcal{B}(B\to K^{(*)}\nu\bar\nu)$. Both implementations of the 4321 model (Model I and II in Table~\ref{tab:VLcontent}) predict 15\% to 60\% enhancement with respect to the SM expectation, in the parameter region which provides a good fit to the B-anomalies. The Belle II Collaboration should be able to measure this branching fraction with a $10\%$ error assuming the SM value, probing all of the parameter space of the 4321 models~\cite{Kou:2018nap}.

\section*{Acknowledgments}

I am grateful to my collaborators Javier Fuentes-Mart\'{\i}n, Gino Isidori and Matthias K\"{o}nig. This project has received funding from the European Research Council (ERC) under the European Union's Horizon 2020 research and innovation programme under grant agreement 833280 (FLAY), and by the Swiss National Science Foundation (SNF) under contract 200021-175940.

\end{document}